# Solar X-ray Monitor (XSM) On-board Chandrayaan-2 Orbiter

M. Shanmugam[1], S. V. Vadawale[1], Arpit R. Patel[1], Hitesh Kumar Adalaja[1], N. P. S. Mithun[1], Tinkal Ladiya[1], Shiv Kumar Goyal[1], Neeraj K. Tiwari[1], Nishant Singh[1], Sushil Kumar[1], Deepak Kumar Painkra[1], Y. B. Acharya[1], Anil Bhardwaj[1], A. K. Hait[2], A. Patinge[2], Abinandhan Kapoor[3], H. N. Suresh Kumar[3], Neeraj Satya[3], Gaurav Saxena[4] and Kalpana Arvind[4]

[1]Physical Research Laboratory, Ahmedabad
[2]Space Application Centre, Ahmedabad
[3]U. R. Rao Satellite Centre, Bengaluru
[4]Laboratory for Electro Optics Systems, Bengaluru
*Email: shansm@prl.res.in*

## Abstract

Solar X-ray Monitor (XSM) is one of the scientific instruments on-board Chandrayaan-2 orbiter. The XSM along with instrument CLASS (Chandra's Large Area Soft x-ray Spectrometer) comprise the remote X-ray fluorescence spectroscopy experiment of Chandrayaan-2 mission with an objective to determine the elemental composition of the lunar surface on a global scale. XSM instrument will measure the solar X-rays in the energy range of 1-15 keV using state-of-the-art Silicon Drift Detector (SDD). The Flight Model (FM) of the XSM payload has been designed, realized and characterized for various operating parameters. XSM provides energy resolution of ~180 eV at 5.9 keV with high time cadence of one second. The X-ray spectra of the Sun observed with XSM will also contribute to the study of solar corona. The detailed description and the performance characteristics of the XSM instrument are presented in this paper.

## Introduction

Remote sensing using X-ray wavelength is an effective way and important tool to study planetary objects[1,2]. Obtaining elemental composition is one of the basic inputs necessary for the scientific study of any planetary body. Global maps of elemental composition help to determine the geo-chemical nature of the planetary surface as well as the history of volcanic and impact activity on such bodies[3]. Such maps can be generated by measuring the characteristic X-ray / gamma-ray lines emitted by various elements due to excitation by solar or cosmic radiation. The technique of remote X-ray Fluorescence spectroscopy is particularly suitable to investigate the elemental composition for the atmosphere-less planetary bodies in the inner solar system due to the abundant availability of the exciting radiation in the form of solar X-rays[1,4-6]. This technique can be used to detect the presence of most of the major rock forming elements, such as Mg, Al, Si, K, Ca, Ti and Fe by measuring their characteristic X-ray lines in the energy range of 1 to 10 keV. Intensities of these characteristic lines can provide quantitative abundances of these elements if the intensity and spectral shape of the incident radiation (solar X-rays) is known accurately. Thus, a typical remote X-ray fluorescence experiment consists of two components, a planetary surface viewing X-ray detector to measure the fluorescent spectra and a Sun viewing X-ray detector to measure incoming solar X-ray spectra[7-9]. The remote X-ray fluorescence spectroscopy experiment on-board Chandrayaan-2 orbiter also consists of two independent payloads – CLASS[10] (Chandra Large Area Soft x-ray Spectrometer), which will measure the fluorescence X-rays emitted by the constituent elements of the lunar surface; and XSM (Solar X-ray Monitor) developed at Physical Research Laboratory (PRL), which will measure the incident solar X-ray spectrum. An overview of the XSM design and its earlier configuration was given by Vadawale et al (2014)[11]. Here we present the final design details of the XSM payload as well as its performance measured during the integrated tests with the Chandrayaan-2 orbiter spacecraft.

## Scientific Objectives

The primary scientific objective of XSM is to provide accurate measurements of variations in intensity and spectral shape of the solar X-rays, which can then be used for quantitative interpretation of the lunar X-ray fluorescence spectra measured by the CLASS experiment. For this purpose, XSM will measure the solar X-ray spectra in the energy range of 1 – 15 keV with energy resolution better than 180 eV (@5.9 keV). These measurements are also very useful for deriving independent scientific information about various high-energy processes occurring in the solar corona. Thus, additional scientific objectives of XSM are to address various outstanding solar physics issues[12], such as – the study of the evolution of non-thermal processes in solar flares, characterization of the elemental abundance in the solar corona and study of X-ray spectra in absence of active regions on the Sun.

## Salient features of XS M

It is well known that the solar X-ray spectrum is highly variable in time depending on the solar activity. During large solar flares, the solar flux can increase by many orders of magnitude within few seconds. Accurate measurement of the solar X-ray flux and its spectrum over such a wide range is a highly challenging task. XSM is designed to measure the incident solar X-ray spectrum in the energy range of 1 to 15 keV with energy resolution better than 180 eV @ 5.9 keV for a wide range of solar X-ray intensity. Table 1 provides an overview of the specifications of the XSM instrument, as well as some of the salient features of XSM, that enables it to accomplish the challenging task, as described later.

| Parameter | Value |
|---|---|
| Energy range | 1 – 15 keV |
| Energy resolution | ~ 180 eV @ 5.9 keV |
| Detector | Silicon Drift Detector (SDD) |
| Detector size | 30 mm$^2$, 450 µm thick |
| Detector aperture | 0.384 mm$^2$ |
| Detector operating temperature | -35º C using inbuilt Peltier cooler |
| Field of view | ±40 degree |
| On-board calibration | $^{55}$Fe source |
| Moving mechanism | 3 stop (open, close, Be-filter) |
| Large flare detection | Automatic/Manual |
| Quantization | 10 bit |
| Cadence | 1 second |
| Dynamic range for one second spectrum | B2 to X5 class flares |

**Table 1: XSM Instrument Specifications.**

**Silicon Drift Detector**

XSM uses the state-of-the-art Silicon Drift Detector (SDD) for measuring the energy of the incident X-ray photons. SDD is functionally similar to a Si-PIN photodiode but has different electrode structure, which results in very low detector capacitance and therefore, under similar operating conditions, the SDD gives better energy resolution than a Si-PIN diode of same size. The low detector capacitance also enables it to maintain the energy resolution at high count rates, a highly desirable feature for a solar X-ray detector. SDDs have been used in space experiments earlier on-board the Mars Rovers[13,14], MinXSS[15] - a cubesat experiment for measuring solar X-ray spectrum and Neutron star Interior Composition Explorer (NICER)[16] for Soft X-ray timing and spectroscopy of astrophysical sources.

**Active temperature control of the detector**

Temperature of all systems would vary substantially during different phases of lunar orbit. Performance of the X-ray detectors depend on their operating temperature. The SDD module used in XSM includes a built-in thermo-electric cooler which is actively controlled to maintain the operating temperature at about -35°C. For XSM, the operating temperature of SDD can also be changed by command if necessary.

**High time cadence**

During solar flares, the X-ray flux can increase by several orders of magnitude within a few seconds and the X-ray spectrum during such flare changes continuously. The X-ray fluorescence spectrum from the lunar surface arising due to fast varying solar X-rays also varies on a similar time scale and thus it is important to measure solar X-ray spectra on the same time scale. XSM is designed to measure entire energy spectrum every second, whenever Sun is within the field of view. It may be noted that this will be the highest cadence measurement of the solar X-rays made so far.

**Field of View**

XSM detector is not designed to track the Sun, rather it is fixed mounted on Chandrayaan-2 Orbiter, which will be orbiting the Moon. Given this constraint, XSM is designed with a large field of view (FOV) of ±40°, so that the Sun can be within the FOV for large fraction of the orbital period. The large FOV of XSM is achieved by using an optimum combination of the active detector area and specially designed detector cap as well as spacecraft mounting bracket.

**Active mechanism**

One of the key requirements from XSM is to obtain the solar X-ray spectrum of all classes of flares without significant degradation in the spectral characteristics. During very large solar flares (class M5 and above) the photon flux rate becomes extremely high resulting in pulse pileup and degraded energy resolution. To avoid this, XSM contains an active mechanism which brings a 250 µm thick Beryllium (Be) filter in the field of view of the detector to reduce the incident photon flux, but temporarily increases the energy threshold to ~2 keV. This maintains the rate of incident photon on SDD within the decided limit of around $1 \times 10^5$ counts/s where the XSM spectral performance is expected to be stable. This is achieved by a real-time flare detection logic which continuously monitors the detector count rate and activates the mechanism when the count-rate crosses the pre-defined threshold value which is commandable. Since the spectrum of larger flares is significantly harder (largest flux at higher energy), the increase in energy threshold does not drastically affect the accurate measurement of the overall energy spectrum. The mechanism also has an additional position housing the onboard calibration source ($^{55}$Fe) which will be used for regular calibration of the instrument.

**High time resolution light-curves**

Apart from providing spectrum at full energy range and high resolution spectrum at every second, XSM is also designed to provide light-curve in three independent energy ranges with cadence of 100 ms. These energy ranges can be chosen anywhere within the overall energy range of XSM and can also be changed through ground command after launch. This feature will be very useful for high time resolution studies of various solar processes.

**XSM Instrument Description**

XSM payload is designed as two separate packages, namely XSM sensor package and XSM processing electronics package. The XSM sensor package houses SDD detector module, analog front-end electronics, high voltage bias generation circuit, and a stepper motor based active threshold control mechanism. XSM processing electronics package consists of Field Programmable Gate Array (FPGA) based payload data readout and control electronics, DC-DC converters, power switching circuits based on telecommand and driver for stepper motor mechanism. The block schematic of the XSM payload is shown in figure 1.

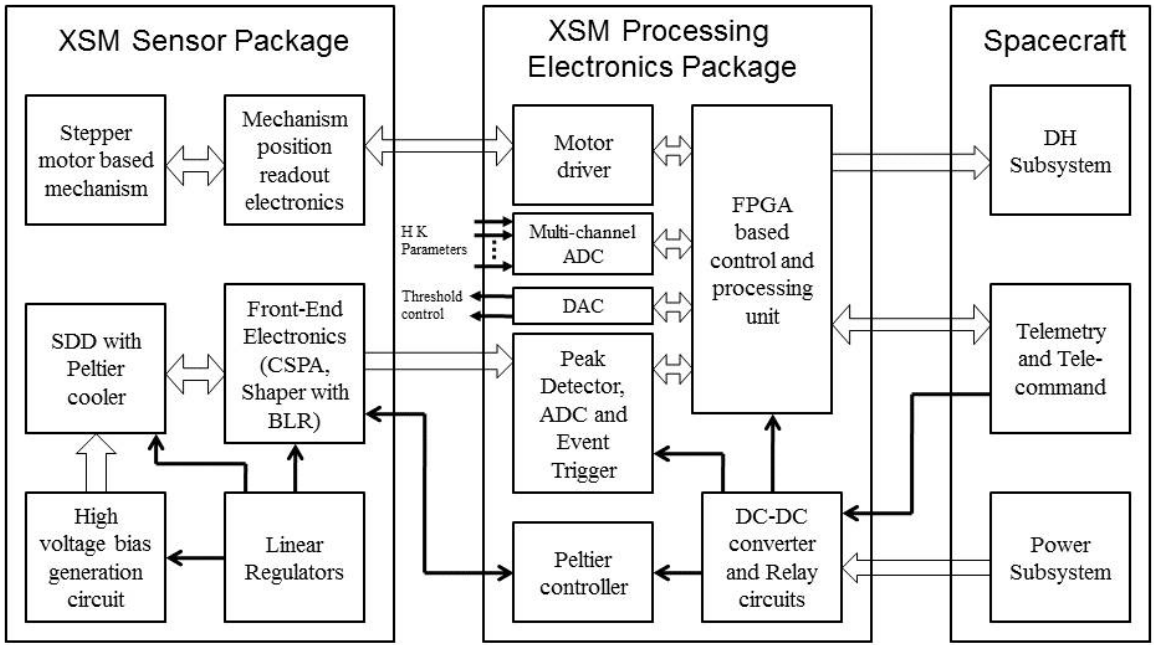

**Fig. 1: Block schematic of XSM payload.**

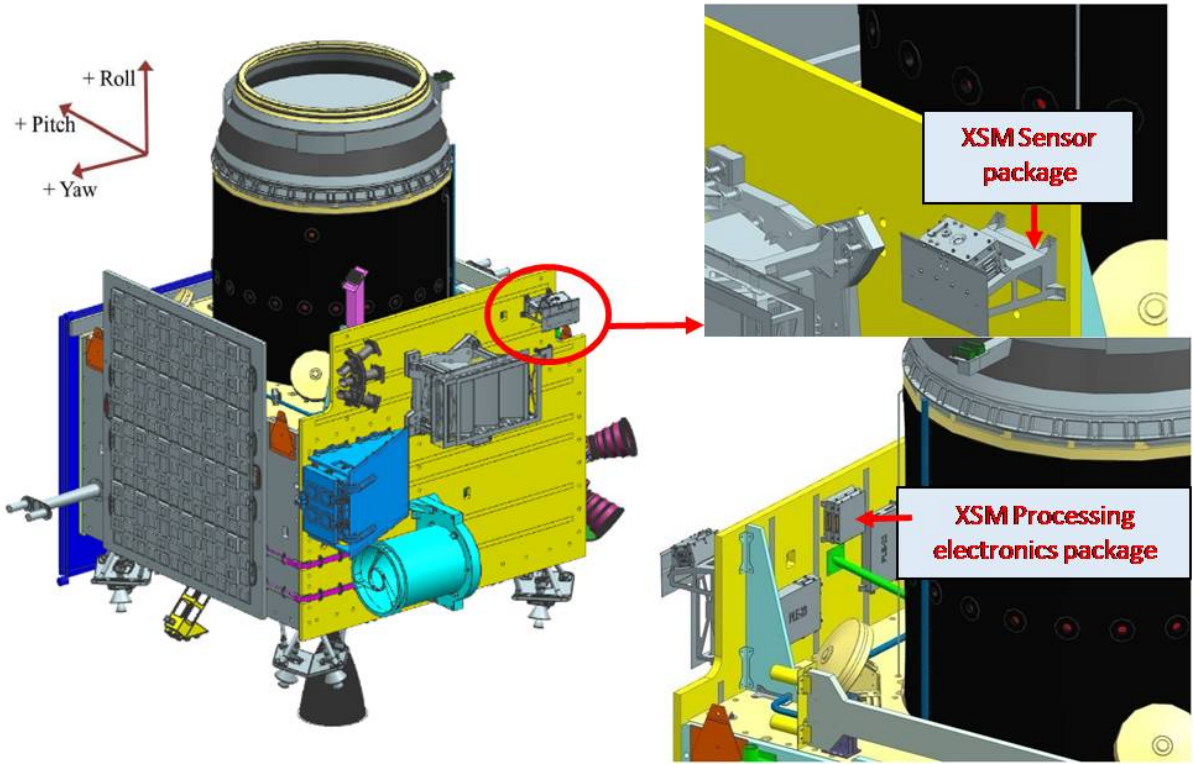

**Fig. 2: Mounting locations of XSM packages on Chandrayaan-2 spacecraft.**

The XSM sensor package is mounted outside the negative pitch panel of the spacecraft and the processing electronics package is mounted inside the negative pitch panel as shown in figure 2. The sensor package is mounted on a bracket with the bore sight of the instrument canted at an angle of 20 degrees from positive roll direction so that all the spacecraft structures clear the field of view of XSM and the radiator plate faces negative pitch direction which is away from Sun.

**XSM sensor package**

XSM sensor package houses the SDD module along with sensitive front-end electronics and the stepper motor based mechanism. The present packaging concept has been arrived based on various functional as well as thermal considerations. The size of the XSM sensor package is 234 x 112 x 133 mm$^3$ and mass is about 650 g. Figure 3 shows the fully assembled XSM sensor package.

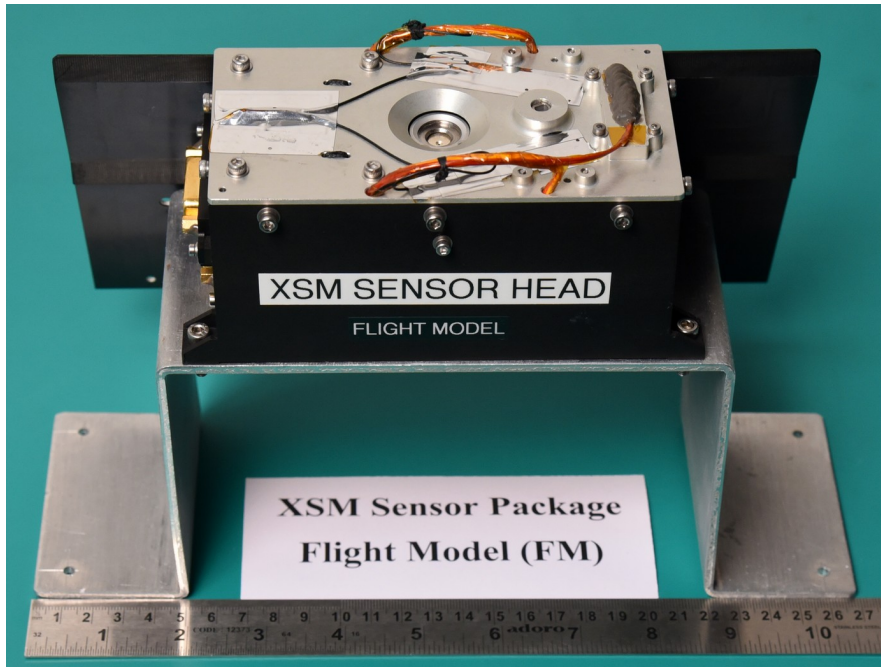

**Fig. 3: XSM Sensor package Flight Model.**

This package contains one Printed Circuit Board (PCB) of size 130 mm x 70 mm accommodating the front-end electronics circuits and another PCB of size 63 mm x 49 mm for motor position sensing and two small PCB strips for mounting the IR Light Emitting Diodes (LED) and IR detectors.

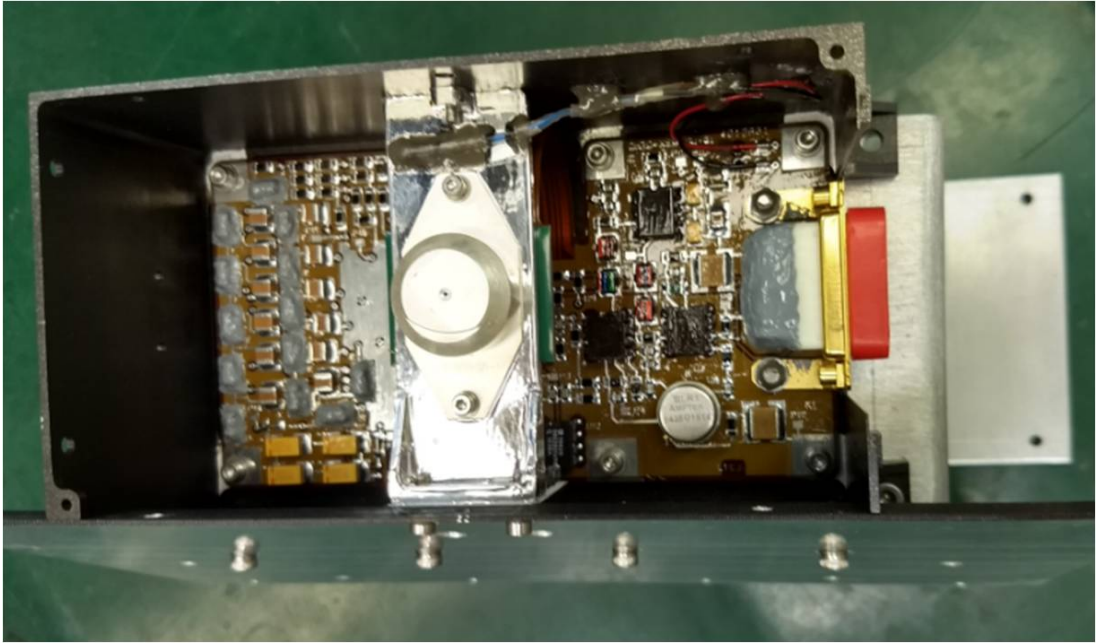

Fig. 4: XSM sensor package without mechanism.

*Detector and Front-End Electronics (FEE)*

Figure 4 shows the XSM sensor package without mechanism, where the detector cap and the front-end electronics PCB are visible. The SDD used in XSM is procured from KETEK, GmBH, which is available in the form of an encapsulated module containing thermo-electric (Peltier) cooler, temperature sensor as well as few critical components of the front-end electronics, such as a Field Effect Transistor (FET), a feedback capacitor and a reset diode. The active area and thickness of the SDD used in the XSM experiment is ~30 mm$^2$ and 450 μm, respectively. SDD module is available in a standard TO-8 package with a 8 μm thick Be window on the top. The in-built peltier element has the capacity to achieve the maximum differential temperature of 75°C. Figure 5 shows the photographic view of the detector module with and without encapsulation.

The SDD module is covered with a detector cap to define the aperture of 0.7 mm diameter. The aperture size is optimized based on the simulations of X-ray spectra expected from different classes of flares to accommodate the solar flares ranging from B to X class. The detector cap is made of 0.5 mm thick Aluminium and is covered with 50 µm Silver coating on both side to block X-rays coming beyond the FOV.

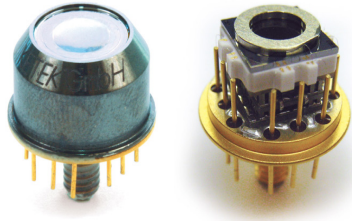

**Fig. 5: Photographic view of SDD module with and without external cover (Courtesy KETEK, GmbH, Germany).**

To achieve the desired performance, SDD is required to be maintained at temperature of around -35°C, which is achieved by closed-loop control of the thermo-electric cooler. The detector is mounted on an Aluminium bracket which is interfaced with a radiator plate for proper thermal management. The electrical interface of the SDD module with the front-end electronics is established through a flexi-rigid interface cable.

XSM Front-End Electronics (FEE) consists of Charge Sensitive Pre-Amplifier (CSPA) that directly interfaces with SDD module, followed by a shaping amplifier with base line restorer. The CSPA converts the charge cloud produced in the SDD due to incident X-ray in to voltage form. The XSM FEE design incorporates "Reset type" CSPA which is compatible with the SDD module used. The output of reset type CSPA is in the form of a ramp signal consisting of charge pulses due to X-ray interaction and also the detector leakage current. X-ray events will appear in the form of steps on the ramp signal and the amplitude of the step pulses depends on the energy of the X-rays interacting with the detector. The ramp signal frequency varies with X-ray energy, incident X-ray rate and the detector temperature. When there is no solar X-ray photon interacting with the detector, the ramp frequency depends on leakage current of the detector. The ramp signal output is discharged before it reaches saturation through a reset pulse generator circuit and a reset diode inside the SDD module. The feedback capacitance which is connected internally has value of ~ 60 fF, which gives the charge to voltage conversion sensitivity of ~ 4 mV for 5 keV X-ray photon.

The output of CSPA is fed to a shaping amplifier which converts the step pulses into semi-Gaussian pulses to achieve good energy resolution. The selection of shaping time constant of the shaping amplifier is an important criterion and needs to be chosen carefully by considering the resultant compromise between the energy resolution and resultant dead time, which limits the count rate handling capability of the instrument. The shaping amplifier consists of three stage amplifier with CR-(RC)$^2$ filters which provide peaking time of pulses about 0.4 μs with gain of about ~60 for covering the energy range of 1-15 keV with the maximum pulse amplitude of 5 V required for the Analog to Digital Converter (ADC). The shaping amplifier also includes a Base Line Restorer (BLR), which minimizes the base line fluctuations during high incident X-ray rates and thereby improving the energy resolution.

Circuit for generation of high voltage biases required for SDD operation is also housed in the sensor package along with the front-end charge readout electronics. The voltage levels of -20 V for inner ring R1, -130 V for outer ring Rx and -60 V for back contact Rb are generated using a voltage multiplier based circuit with an input voltage of 12 V.

*Stepper motor based mechanism*

In order to avoid saturation during high intensity flares while being sensitive for the low intensity ones, an innovative stepper motor based mechanism with Be filter has been developed for the XSM payload. This mechanism has a filter wheel with three positions corresponding to Open, Calibration source, and Be filter, mounted on the shaft of the stepper motor. The filter wheel also has a provision to indicate its position using IR LEDs and sensors. The output of these IR sensors are used by the processing electronics to identify the current position and also used as a feedback for the subsequent mechanism movement.

The on-board calibration source mounted on the mechanism is $^{55}$Fe with activity of ~100 µCi. It is also covered with 3 µm Titanium foil producing total four X-ray lines (Mn Kα, Kβ and Ti Kα, Kβ with energies of 5.89, 6.49, 4.51 and 4.93 keV respectively) for accurate energy calibration. The source is custom made by BARC for use in XSM. The calibration source and the Be window is fixed in the filter wheel as shown in figure 6. The mechanism assembly forms the top cover of the sensor package.

The mechanism is designed to operate in three modes namely automatic mode, manual mode and force-step mode. In automatic mode, the mechanism moves from open position to Be filter position when the incident count rate is above a threshold value, that can be set with data command, for consecutive five 100 ms intervals. This is expected during M/X class solar flares. When the count rate with Be filter is below another commandable threshold rate, the mechanism moves back to open position. In order to avoid the mechanism to move back and forth in the case of count rate being close to threshold values, minimum duration of one minute is set between successive automated movements of the mechanism. In manual mode, the motor movement is controlled by ground command, allowing movement between any of the positions of the filter wheel. Manual mode will be used for XSM on-board calibration. As a contingency plan in case of failure of position sensing LEDs or detectors, a force-step mode is also implemented in which motor can be moved one step at a time based on ground command.

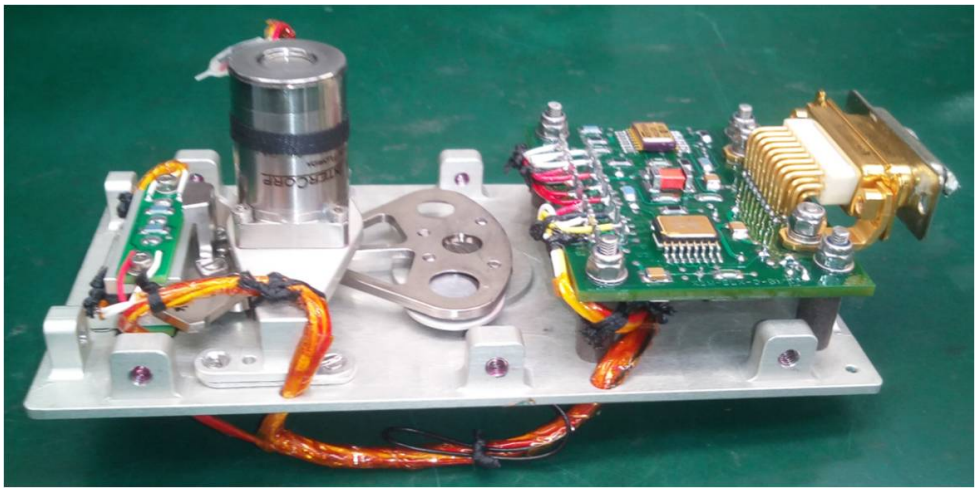

**Fig. 6: XSM mechanism.**

**XSM processing electronics package**

XSM processing electronics package is responsible for generating required voltages from raw power, controlling the XSM sensor package, as well as all on-board processing of the real time data and communicating with spacecraft. The size of the XSM processing electronics package is 150 x 127 x 55 mm$^3$ and mass is about 700 g. Flight model of XSM processing electronics package is shown in figure 7.

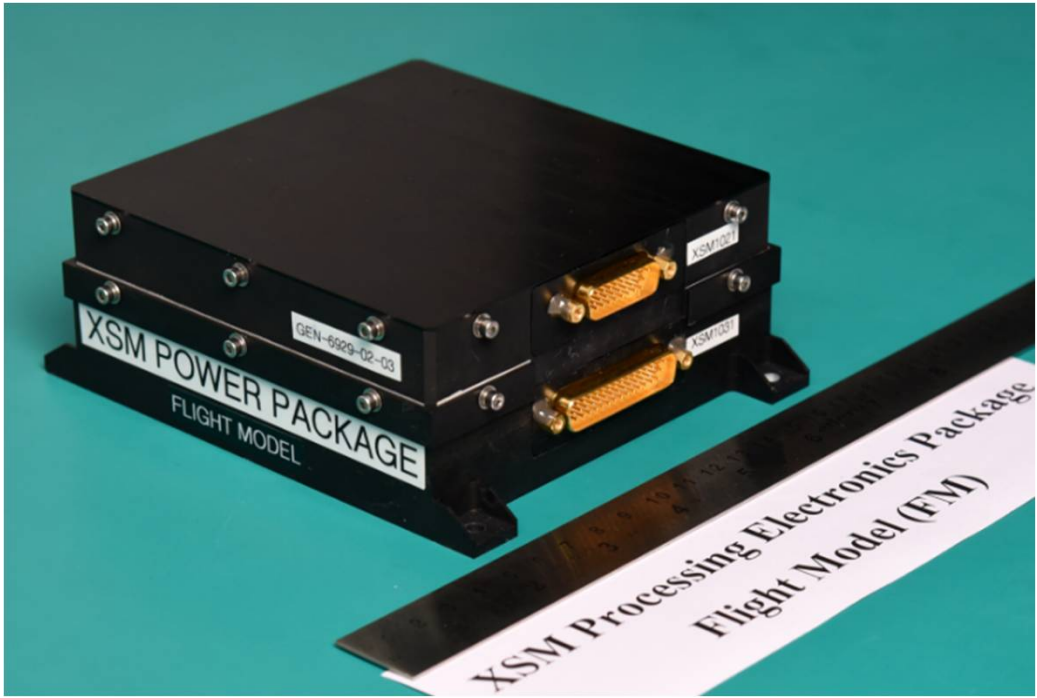

**Fig. 7: XSM Processing Electronics Package Flight Model.**

XSM processing electronics package consists of a stack of two PCBs. The FPGA PCB (housed on the top tray) shown in figure 8 contains event trigger generator, peak detector, ADC and FPGA based real time data acquisition & payload control and interface with Chandrayaan-2. The power PCB (housed on the bottom tray) contains DC-DC converters, power ON/OFF circuits, Peltier controller and the stepper motor driver for XSM mechanism.

The FPGA PCB receives the analog signal output of the shaping amplifier from XSM sensor package which is then fed to event trigger generator and the peak detector. FPGA initiates to hold the peak amplitude of analog pulses on occurrence of the event trigger signal i.e. whenever the shaping amplifier output crosses the pre-defined threshold level. The stored peak amplitude is then converted into digital form using A/D converter. XSM uses a 12 bit serial ADC having conversion time of ~ 880 ns. FPGA reads the 10 MSB bits of event data and updates the 1024 channel spectrum which is stored in the internal memory. This process repeats for each X-ray interaction for the duration of one second and at the end of one second, FPGA adds the header data into the spectral data to form a data packet of 1024 x 16 bits. FPGA uses two banks of 1024 x 16 bits memory locations for storing the spectral and header data in the internal memory. At any given point of time, one bank of memory stores the spectral data for one second and at the same time, data stored in the second bank is transmitted to the spacecraft data handling system. This provides solar X-ray spectrum at every second and also results in a fixed data rate of 2 kB per second irrespective of the incident count rate. The header data consists of health parameters associated with the payload. The header also includes light curve with 100 ms time resolution for three defined energy ranges. These energy ranges can be modified with data commands.

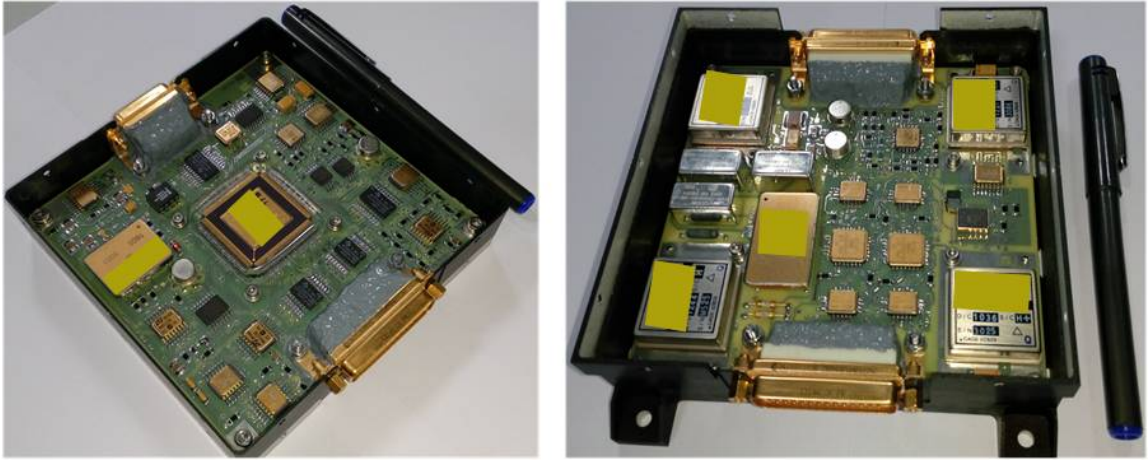

**Fig. 8: FPGA PCB (left), Power PCB (right) of XSM processing electronics.**

XSM power PCB receives raw power (28-42 V) from the spacecraft power system and all the required bias voltages for the XSM payload operation are generated internally using DC-DC converters and Low Drop Out (LDO) regulators. These converters are low power converters available in smaller size, PCB mountable and lesser weight.

This package also includes the circuit for active control of the Peltier cooler internal to the SDD module for maintaining the detector temperature. The thermistor inside the SDD module is used as feedback for the closed loop control. This is realized by controlling the gate voltage of N-Channel FET and thereby the Peltier current, with respect to the difference between the set point and the measured temperature with the thermistor. Maximum duration of two minutes is required for achieving the nominal detector operating temperature of -35°C from ambient conditions of ~25°C. In case of requirement of maintaining the detector at temperatures other than -35°C, there is also provision to set the desired temperature value with data command.

## Instrument performance

The Flight Model of XSM is realized after verification of performance of the qualification model during various environmental tests. Further, the flight model of the instrument has undergone extensive test and evaluation. The $^{55}$Fe calibration source with titanium foil which is included in the instrument is used to evaluate the performance of XSM during the tests.

Figure 9 shows the spectrum acquired with XSM with the calibration source in the mechanism brought in front of the detector, during the nominal operation of the instrument with the detector cooled to -35°C. X-ray lines from the source are detected in the spectrum and the energy resolution of the instrument, measured as the full width at half maximum (FWHM) of the Gaussian line at 5.89 keV, is shown to be ~175 eV, which is better than the desired specification of 200 eV. The energy resolution of the instrument after being exposed to environmental conditions was same as that of pre-test results within the accuracy of ±2 eV.

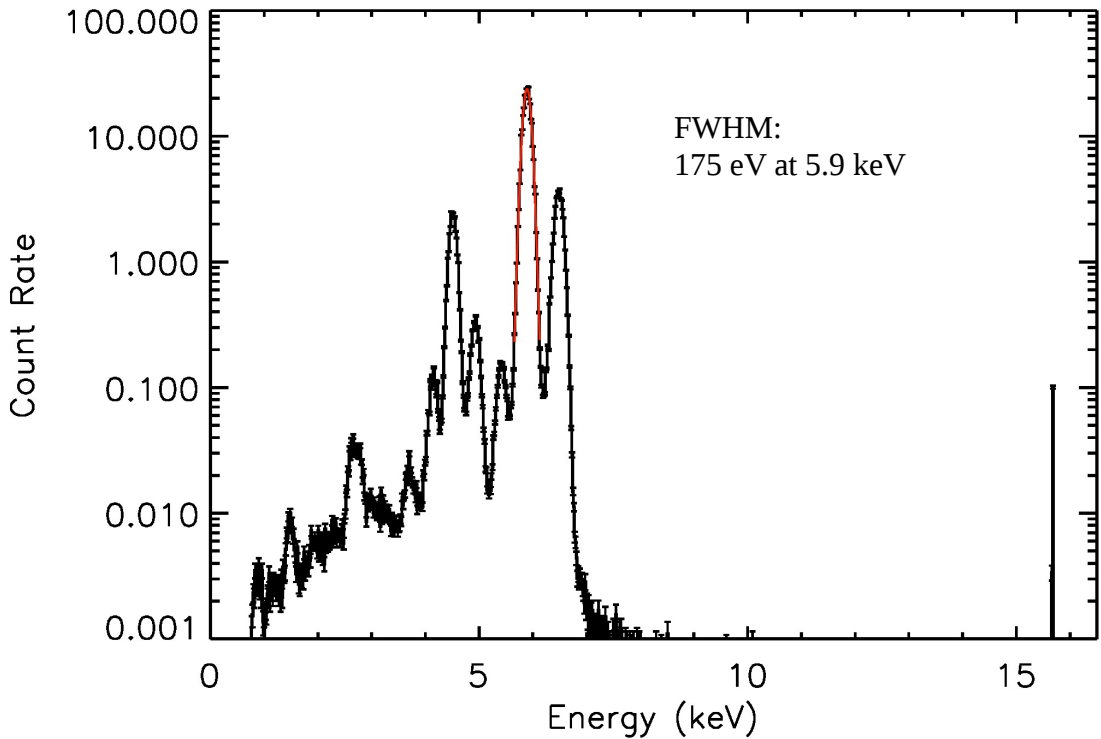

**Fig. 9: XSM spectrum acquired with calibration source. Lines at 5.89, 6.49, 4.51 and 4.95 keV are prominent along with other fainter lines in the spectrum. The spectral resolution measured as FWHM of line at 5.89 keV is 175 eV.**

It is well known that in X-ray spectrometers, the peak energy channel varies with the incident rate and the energy resolution (FWHM) is expected to degrade beyond a limit of count rate. In order to verify the performance of XSM as a function of incident rate, an experiment was carried out at X-ray beam line BL-16 of Indus-2 synchrotron facility at RRCAT, Indore. Data was acquired with XSM by varying the incident rate using slits in the beam line. Monochromatic line at 5.0 keV is fitted with a Gaussian to obtain peak channel and FWHM as a function of incident rate as shown in figure 10. It is to be noted that up to an incident rate of $1 \times 10^5$ counts/s, the variation in peak channel is less than

~2% and FWHM variation is within 4%. Beyond this rate, both peak channel and FWHM begin to vary drastically. This shows that the spectral performance of XSM is not affected significantly up to incident count rates of $1\times10^5$ counts/s, which corresponds to typical count rate for M5 class flare without Be filter and X6 class flare with Be filter.

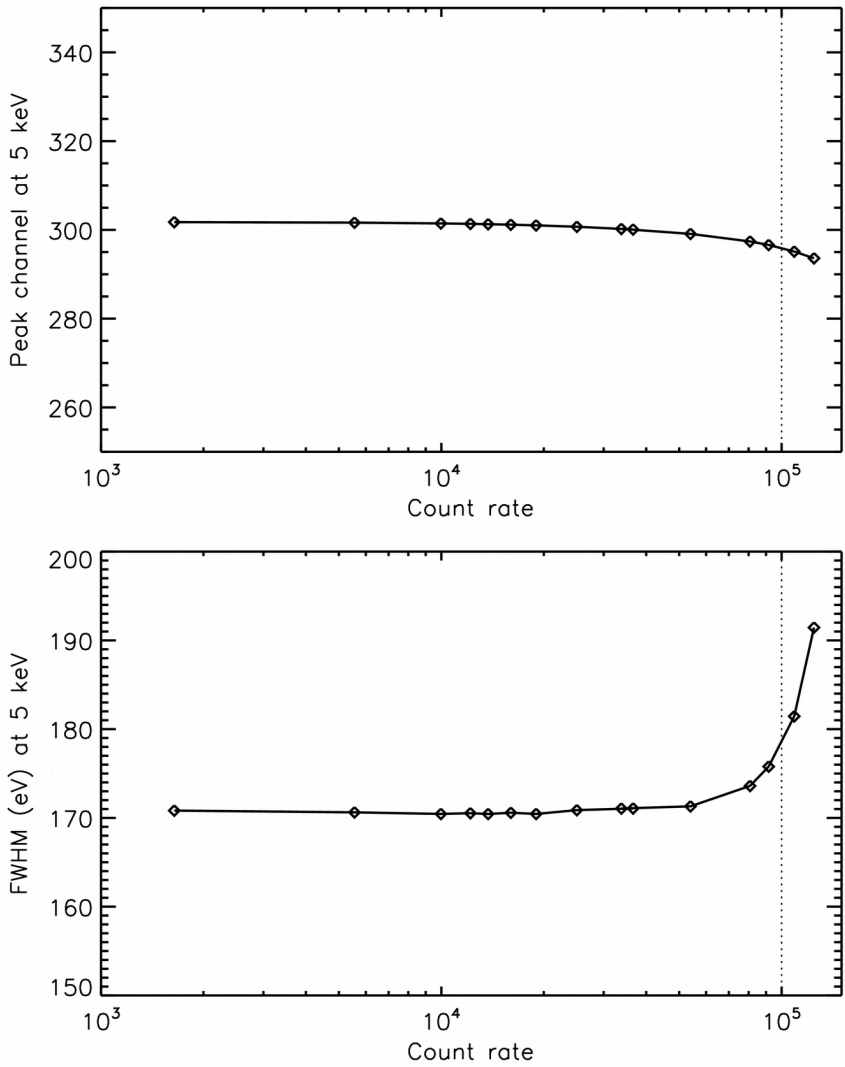

**Fig. 10: Peak channel and FWHM variation with incident count rate at 5 keV. The spectral performance does not vary significantly till $1\times10^5$ counts/s.**

To obtain the incident Solar X-ray spectrum from the observations, it is required to accurately model the response matrix of the instrument. Extensive calibration experiments are carried out to obtain gain correction factors, redistribution matrix parameters, Be filter transmission, detector efficiency, dead time characteristics, pileup, detector collimator angular response etc, results of which will be reported elsewhere.

The XSM instrument is integrated with the orbiter craft of Chandrayaan-2 and has undergone various integrated spacecraft level tests where the instrument performance is measured using the $^{55}$Fe calibration source included in the payload. The obtained spectra are identical to that shown in figure 9, and hence the performance of the instrument is as per specifications.

**In-orbit operations and data analysis**

After in-orbit commissioning phase, XSM is expected to carry out solar observations whenever Sun is within the field of view of the instrument and provide solar flux measurement every second. As XSM is fix mounted on the spacecraft, the duration when Sun is within the field of view would depend on the inertial orientation of the orbit. During noon-midnight orbits, the typical observation duration would be ~20 minutes of every ~110 minute orbit, whereas in dawn-dusk orbits, continuous observations would be available. It is also planned to carry out regular calibration of XSM with the on-board calibration source.

Basic XSM data will be available in the form of full energy spectrum recorded every second. The data from Chandrayaan-2 spacecraft will be downloaded at Indian Space Science Data Center (ISSDC), Bylalu, where payload-wise segregation and level-0 processing will also be carried out. The level-0 data will be sent to XSM Payload Operations Center (POC) at Physical Research Laboratory (PRL), Ahmedabad, where higher levels of data products will be generated using XSM Data Analysis Software (XSMDAS). The basic data product from XSM is a time series of calibrated Solar X-ray spectra in FITS format, compatible with standard X-ray spectral analysis tools. Daily plots of light curves of Solar X-ray emission as measured by XSM will be made available online on a regular basis. Data will be available publicly for download from ISSDC after specified lock-in period.

## Summary

The flight model of the XSM payload has been developed and characterized for performance requirement and also tested for various environmental conditions. XSM payload meets all the design criteria and shown to provide the energy resolution of ~ 175 eV at 5.9 keV with low energy threshold of ~800 eV. XSM will provide the highest cadence measurements of solar X-ray spectra available till date. It has been integrated with Chandrayaan-2 Orbiter and tested at various phases, including environmental tests. Performance of the instrument is normal during all these tests.

## Acknowledgements

XSM payload is designed and developed by Physical Research Laboratory (PRL), Ahmedabad, supported by Department of Space. PRL is also responsible for the development of data processing software, overall payload operations and data analysis of XSM.  The active mechanism for XSM is provided by U. R. Rao Satellite Centre (URSC), Bengaluru along with Laboratory for Electro Optics Systems (LEOS), Bengaluru. Thermal design and analysis of XSM packages was carried out by URSC whereas Space Application Centre (SAC) supported in the mechanical design and analysis. SAC also supported in the fabrication of the flight model of the payload and its test & evaluation for the flight use. Authors would like to thank various facilities and the technical teams of all above centers for their support during the design, fabrication and testing of this payload. Authors also would like to thank the RRCAT for providing the Indus-2 synchrotron facility for the calibration of the XSM instrument. The Chandrayaan-2 project is funded by Indian Space Research Organization (ISRO).